\documentclass[12pt]{article}
\usepackage[dvips]{graphicx}
\begin{document}
\title{\bf Nucleosynthesis in slowly evolving Cosmologies}
\author{\sf{Pranav Kumar}\footnote{e-mail: pranav@physics.du.ac.in} \ \& 
        \sf{Daksh Lohiya}\footnote{e-mail: dlohiya@iucaa.ernet.in}\\ 
         Department of Physics \& Astrophysics, University of Delhi\\
         Delhi 110 007, India\\}

\date{}

\maketitle      
\begin{center}
Abstract\\
\end{center}
We explore aspects of Cosmological Nucleosynthesis in an FRW
universe in which the scale factor evolves linearly with time: 
$a(t) \sim t$. A high Lepton number density during the period when
significant nucleosynthesis takes place would lead to a dominant 
screening of the Coulomb potential of colliding nucleii. This would 
lead to a significant enhancement of nucleosynthesis rates. We 
demonstrate how adequate amount of $^4He$ and a collataral 
metallicity, close to the lowest 
metallicity observed in metal poor Pop II stars and clouds, can be 
produced with such an evolution. 

\section{Introduction:}

In an earlier article \cite{dl1}
we reviewed Standard Big-Bang Nucleosynthesis [SBBN] and  
reported a study of nucleosynthesis in a universe 
in which the scale factor evolves linearly with time independent 
of the equation of state of matter. A strictly linear evolution 
of the cosmological scale factor
is surprisingly an excellent fit to a host of cosmological 
observations. Any model that can support such a coasting presents 
itself as a falsifiable model as far as classical cosmological 
tests are concerned as it exhibits distinguishable and verifiable 
features. The motivation for such an endeavor has been discussed at 
length in a series of earlier articles \cite{meetu,abha}. Linear 
Coasting turns out to be broadly concordant with Classical Cosmology 
tests. What makes linear coasting particularly appealing is a 
straightforward adaptation of standard nucleosynthesis codes to
demonstrate that primordial nucleosynthesis is not an impediment 
for a linear coasting cosmology \cite{annu,kapl}. A linear 
evolution of the scale factor radically effects nucleosynthesis
in the early universe. With the present age of the universe some 
$15\times 10^9$ years and the $effective$ CMB temperature 2.73 K, 
the universe turns out to be some 45 years old at $10^9$ K.  
With the universe expanding at such low rates, weak interactions 
remain in equillibrium for temperature as low as $\approx 10^8$ K.
The neutron to proton ratio is determined by the n-p
mass difference and is approximately $n/p\sim exp[-15/T_9]$ as long 
as weak interactions are in equilibrium. This falls to abysmally low 
values at temperatures below $10^9$ K.
Significant nucleosynthesis leading to helium formation commences 
only near temperatures below $\sim 5\times 10^9$K. The low n/p
ratio is not an impediment to adequate helium production. This
is because once nucleosynthesis commences, inverse beta decay, which
does not frreze at these temperatures,
replenishes neutrons by converting protons into neutrons and 
pumping them into the nucleosynthesis network. For baryon entropy 
ratio $\eta\approx 7.8\times 10^{-9}$, the standard 
nucleosynthesis network can be modified for linear coasting and gives 
$\approx 23.9\% $ Helium.
The temperatures are high enough to cause helium to burn.
Even in SBBN the temperatures are high enough for helium to burn.
However, the universe expands very rapidly in SBBN. In comparison,
the linear evolution gives enough time for successive burning 
of helium, carbon and oxygen. The metallicity yield is some $10^8$ 
times the metallicity 
produced in the early universe in the SBBN. The metallicity 
is expected to get distributed amongst nucleii with maximum 
binding energies per nucleon. These are nuclei with atomic 
masses between 50 and 60. This metallicity is close to
that seen in lowest metallicity objects. The 
metallicity concommitantly produced with $\approx 23.9\%$ 
Helium is roughly $\approx 10^{-5}$ solar. 

The only problem that one has to contend with is the significantly
low residual deuterium in such an evolution. The desired amount
would have to be produced by the spallation processes much later 
in the history of the universe. In \cite{dl1} we 
demonstrated how observed abundances of light 
elements besides $^4He$ could be produced by spallation reactions in 
incipient Pop II stellar environments without a collateral overproduction
of $^7Li$. It was demonstrated that the
absence or deficiency of heavy nuclei in a target cloud and 
deficiency of alpha particles in the incident beam would clearly 
suppress lithium production in typical spallation reactions. 
Such conditions are expected of the environments of incipient Pop II stars
and easily circumvent the ``no-go'' concern of Epstein et al 
\cite{eps} related overproduction of $^7Li$ associated with
collateral production of deuterium upto observed levels.

In SBB, hardly any metallicity is produced in the very early 
universe. Metal enrichment is supposed to be facilitated by a 
generation of Pop III stars. Pop III star formation from a 
pristine material is not well understood till date, in spite of a 
lot of effort that has been expanded to that effect recently 
\cite{sneider}. It is believed that with metallicity below a 
critical transition metallicity 
($Z_{cr} \approx 10^{-4} Z_\odot$), masses of Pop III stars would 
be biased towards very high masses. Metal content higher than 
$Z_{cr}$ facilitates cooling and a formation of lower mass Pop II 
stars. In SBB, the route to Deuterium by spallation discussed in 
this article would have to follow a low metal contamination by a 
generation of Pop III stars. In SBB, large-scale production 
and recycling of metals through 
exploding early generation Pop III stars leads to verifiable 
observational constraints. Such stars would be visible as 
27 - 29 magnitude stars appearing any time in every square 
arc-minute of the sky. Serious doubts have been expressed on the 
existence and detection of such signals \cite{escude}. The linear 
coasting cosmology would do away with the requirement of such 
Pop III stars altogether.

Unfortunately, the baryon entropy ratio required for the right 
amount of helium in linear coasting cosmology corresponds 
to $\Omega_b \equiv \rho_b/\rho_c = 8\pi G \rho_b/3H_o^2 
\approx 0.69$. This over - closes dynamic mass 
estimates by a little over a factor of two. This article demonstrates 
an easy and physically acceptable manner that one could get the 
observed amount of Helium with a baryon density significantly lower 
than $\Omega_b \approx 0.69$. 

\section{Coulomb Screening}

In stellar nucleosynthesis, the effect of a screening charged cloud 
on the rate of thermonuclaer reactions was investigated by 
Salpeter and others \cite{sal,sch,kel}. The electrostatic potential 
of a bare nucleus induces a spherically symmetric polarization of the 
surrounding electrons and nucleii. Under typical conditions for the 
interior of main sequence stars, this screening leads to an increase 
of thermonuclear reaction rates upto a factor of two. The effect of 
screening is that the reaction rates between interacting nucleii is 
enhanced by a factor $exp(-U_0/kT)$, with $U_0$ the (negative) potential 
of the gas cloud at the origin (the location of the interacting nucleii). 
Although \cite{sal,sch,kel} give expressions for $U_0$ under varying 
conditions, it would be fair to say that there is no rigorous theory of 
screening to date. These papers are at best schematic and establish the 
effect in principle. 

\begin{figure}
\centering
\includegraphics[angle=-90,width=11cm]{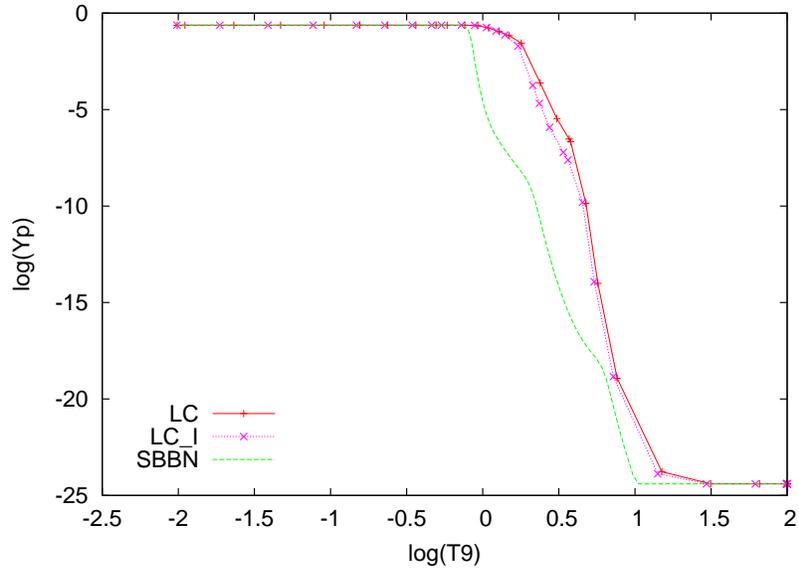}
\caption{Evaluation of helium abundance with temperature is shown for three
cases:(1) Linear Coasting Cosmology with $\eta=7.80\times10^{-9}$ (LC),
(2) Linear coasting model with $\eta=3.159\times10^{-9}$ with the
reaction rate of $D[p,\gamma]^3He$ enhanced by a factor of $6.7$ (LC\_I)
(3) Standard Big Bang Nucleosynthesis (SBBN). }\label{t9he}
\end{figure}
\begin{figure}
\centering
\includegraphics[angle=-90,width=10cm]{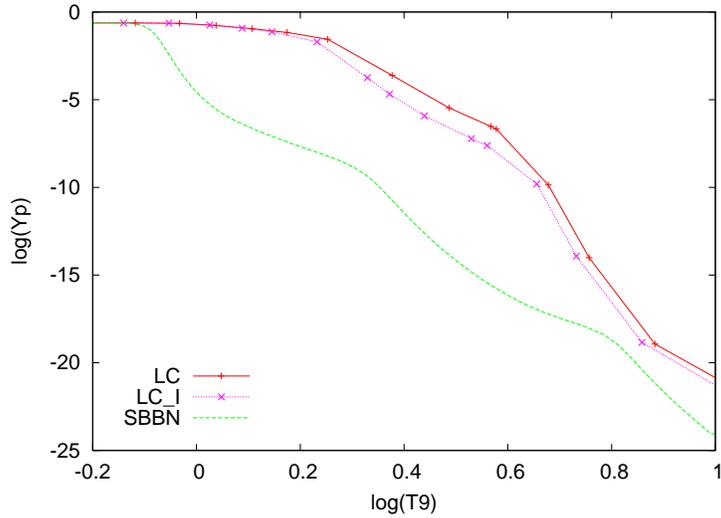}
\caption{The figure shows the temperatue range in which significant helium 
formation occurs.}\label{t9he1}
\end{figure}

Under conditions expected to prevail in a linear coasting cosmology, the
result of a typical run is reported in first two columns of Table 
\ref{table}. As seen in figures \ref{t9he} \& \ref{t9he1}, significant 
nucleosynthesis takes place at temperatures between approximataly 
$8\times10^9K$ and $10^9K$.   

\begin{figure}
\centering
\includegraphics[angle=-90, width=12cm]{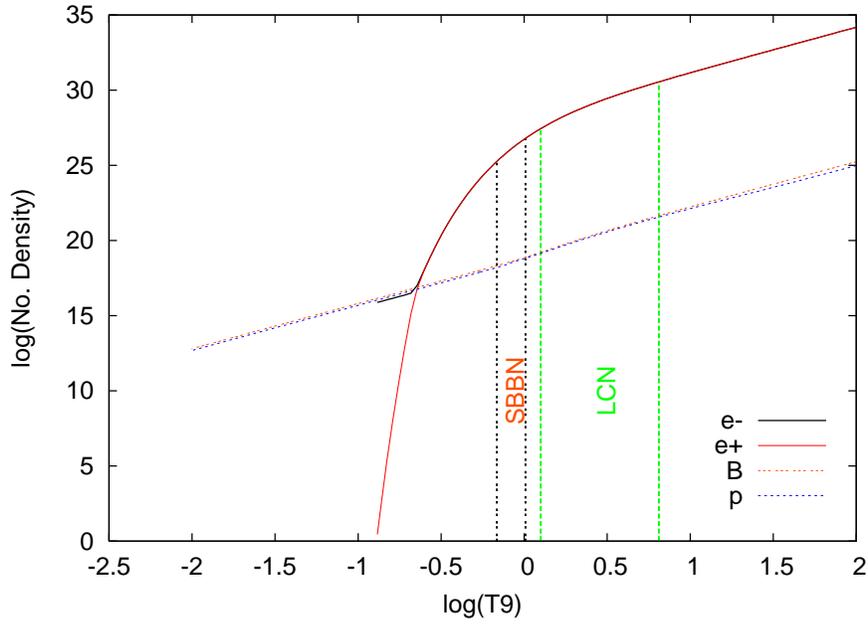}
\caption{Evolution of Number density of electrons, positrons, 
baryons and protons. The verticle lines mark the range of most 
effective nucleosynthesis in SBBN and LCN (Linear Coasting
Nucleosynthesis).}\label{epbp}
\end{figure}

\begin{figure}
\centering
\includegraphics[angle=-90, width=12cm]{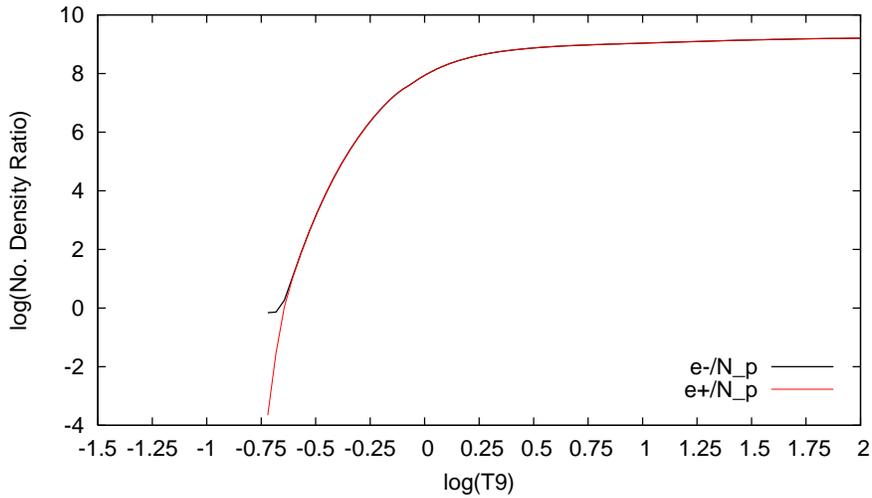}
\caption{Ratio of Number density of $e^-$ to $p$ and $e^+$ to $p$. 
The electron number density drops from $10^{10}$ times proton 
number density at $T\sim10^{11}K$ to equal it around 
$T\sim 2.2\times10^{8}K$.}\label{ratio}
\end{figure}

Figure \ref{epbp} shows the variation of number density of electrons, 
positrons, baryons and protons with temperature. Figure \ref{ratio} shows 
the variation of the ratio of electron to proton and positron to proton 
number densities with temperature. The electron number density drops from 
$\approx 10^{10}$ times the proton number density at $10^{10}K$ to 1 as 
temperature falls below the electron positron annihilation temperature 
of $\approx 10^9K$. The electron - positron cloud would therefore
provide for a very strong screening of the Coulomb potential of the 
interacting nuclei. 

\begin{figure}
\centering
\includegraphics[angle=-90,width=12cm]{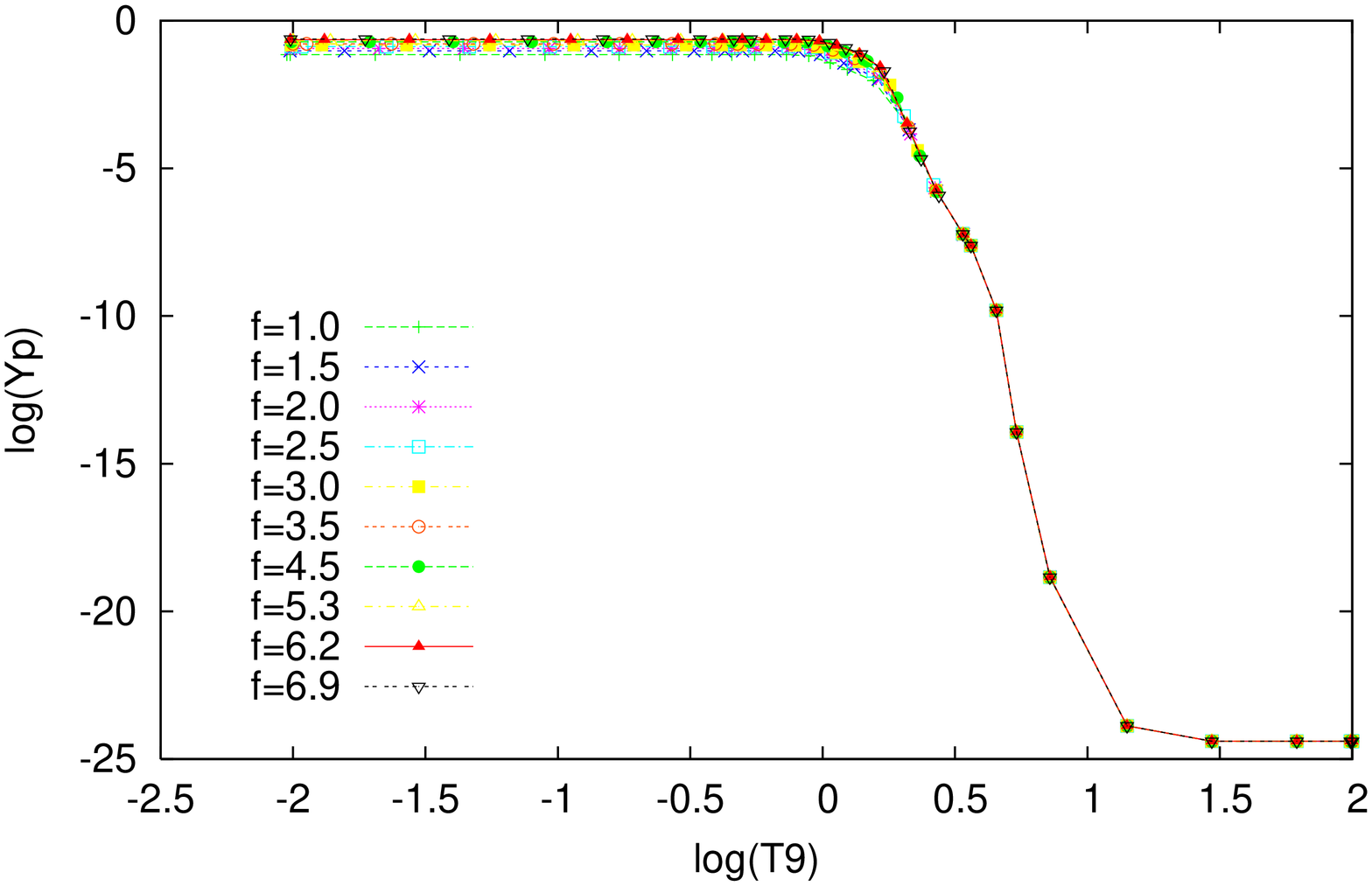}
\includegraphics[angle=-90,width=11cm]{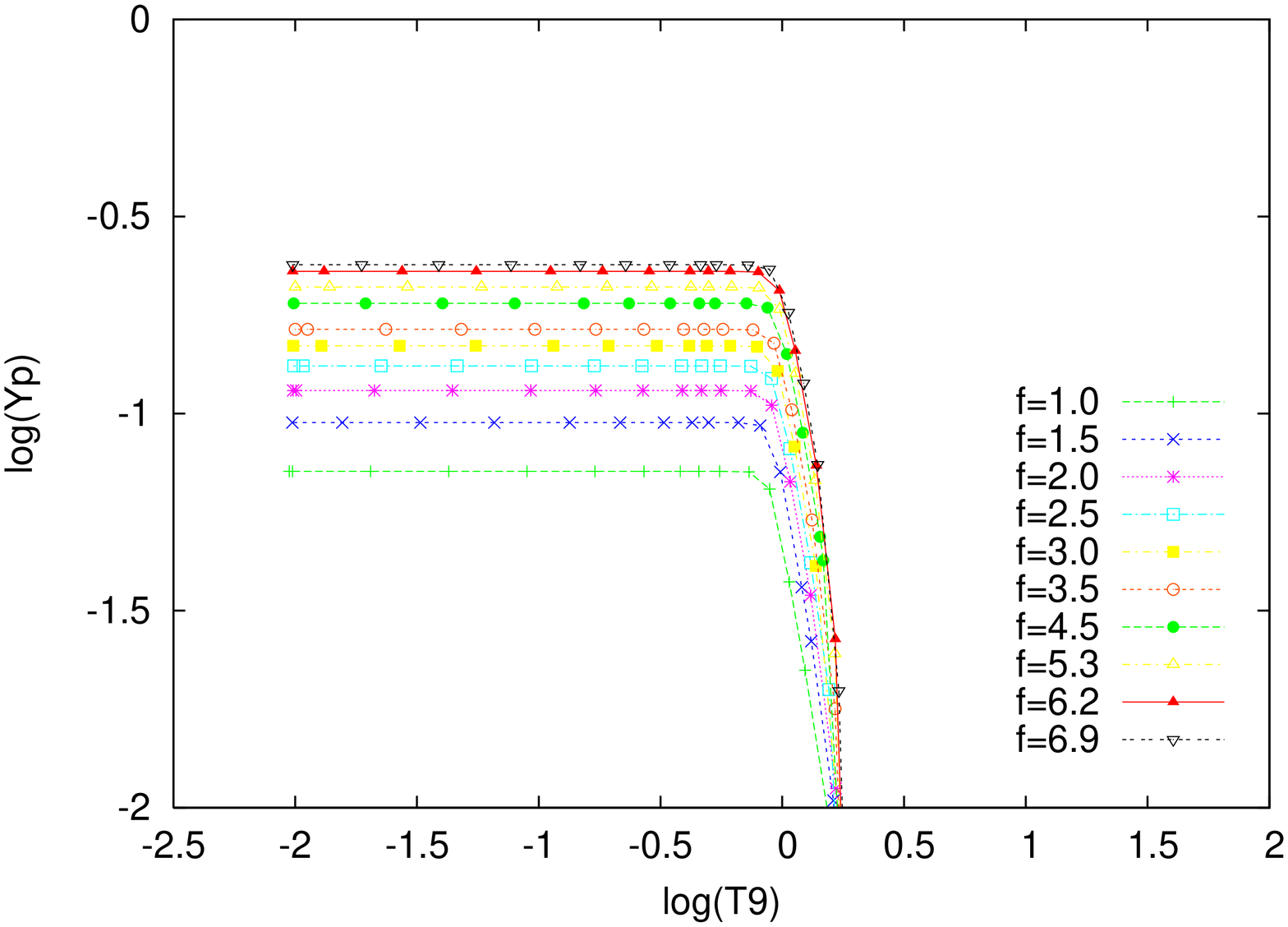}
\caption{Evolution of helium with $T_9$ in Coasting 
Cosmology for $\eta = 3.159\times10^{-9}$ with reaction rate for
$D[p,\gamma]^3He$ enhancing by varying factors `f'.}\label{fchange}
\end{figure}

As one is at a loss for a precise theory of screening, we were satisfied by 
playing around with the nucleosynthesis code and discovered that, among the 
reactions: $^3He+n \rightarrow p+^3H$, $^3H+p \rightarrow \gamma+^4He$, 
$^3H+D \rightarrow n+^4He$, $^3He+\alpha \rightarrow \gamma+^7Be$,
$^7Be+n \rightarrow \alpha+^4He$, $^7Be+D \rightarrow p+\alpha+^4He$,
$^3He+^3He\rightarrow 2p+^4He$, $^2H+p \rightarrow \gamma+^3He$, which
take part in the production of helium, the most effective reaction is 
\begin{equation}\label{f20}
	 ^2H + p \rightarrow ^3He + \gamma. 
\end{equation}
An enhancement of rate of this rection enhances the production of helium. 
The last two columns of Table \ref{table} describes the result of the 
runs for Linear Coasting Cosmology. An enhancement of reaction rate of 
$D [p,\gamma] ^3He$ by a factor of $6.7$ gives the right amount of 
helium ($^4He$) for $\Omega_b = 0.28$ (i.e. $\eta = 3.159\times10^{-9}$).  
Figure \ref{fchange} shows the variation of abundances of helium 
with the change in enhancement factor for the same value of $\eta$. 
In figures \ref{t9he} \& \ref{t9he1} the curve labeled ``LC\_I'' describes 
the profile of helium abundance with temperature in this case. 
In \cite{sal}, for weak screening, the expression for the enhancement 
factor is given by $exp(-\frac{U_0}{kT})$ with 
$$-\frac{U_0}{kT} = \frac{Z_1Z_2e^2}{R(kT)}$$
where $R$ is the radius of charge cloud. An enhancemene of this by a 
factor of $6.7$ for $D [p,\gamma] ^3He$, i.e.
for $Z_1=Z_2=1$, implies an enhancement 
$(6.7)^{Z_1Z_2}$ for higher charged nucleii. This would lead
to a significant creation of metallicity. However, the condition of 
weak screening no longer holds and we have therefore been content to 
report the plausiblity of such a possibility.

We conclude that in principal it is possible to produce all light elements and 
metallicity seen in lowest metallicity objects in a linear coasting cosmology.
With evolving electron, positron and proton number densities in the early 
universe, the screening enhancement factor would in general evolve with 
temperature. The purpose of this article is to make out a case for a better
theory. Definitive results would have to therefore await a rigorous theory
of screening that would play a vital role in such slowly evolving models.

\vspace{0.2cm}
\vfil
\noindent{\large \bf Acknowledgment.}\\ 
Daksh Lohiya and Pranav Kumar acknowledge University Grants Commission and Council of Scientific and Industrial Research, Government of India for financial support.


\bibliography{plain}

\begin {thebibliography}{100}

\bibitem{dl1}G. Sethi, P. Kumar, S. Pandey and D. Lohiya; asrto-ph/0502370
\bibitem{meetu}M. Sethi, A. Batra \& D. Lohiya; $Phy.~Rev.~ \underline {D60}$, 3678 (1987).
\bibitem{abha}A. Dev, M. Safanova, D. Jain \& D. Lohiya; $Phy.~ Lett~\underline {B548}$, 12 (2002).
\bibitem{annu} A Batra, D Lohiya S Mahajan \& A. Mukherjee; $ Int$.
$ J.~ Mod.~ Phys.~ \underline{D6} $, 757, 2000.
\bibitem{kapl} M. Kaplinghat, G. Steigman, I. Tkachev,
\& T. P. Walker; $ Phys$. $Rev.~\underline{D59}$, 043514, 1999.
\bibitem{eps}R. I. Epstein, J. M. Lattimer \& D. N. Schramm, Nature
$\underline{263}$, 198 (1976)
\bibitem{sneider} E. Scannapieco, R. Schnieder \& A. Ferrara; astro-ph/0301628.
\bibitem{escude}J. M. Escude \& M. J. Rees; astro-ph/9701093.
\bibitem{sal}E. E. Salpeter; $Au. J. Ph. \underline{7}$, 373 (1954).
\bibitem{sch}E. Schatzman; $J. Phys. Radium \underline{9}$, 46 (1948); $Astrophys. J. \underline{119}$, 464 (1954).
\bibitem{kel} G. Keller; $ Astrophys. J. \underline{118}$, 142 (1953). 

\end {thebibliography}



\begin{table}\caption{\em Evolution of helium with temperature in 
Linear Coasting Cosmology: (1) without screening; $\eta = 7.80\times10^{-9}$ 
and (2) with screening; $\eta = 3.159\times10^{-9}$ and rate of 
$D[p,\gamma]^3He$ enhanced by factor $6.7$.}
\begin{center}
\begin{tabular}{|c|c||c|c|}\hline
\multicolumn{2}{|c||}{\it Linear Coasting without screening}&
\multicolumn{2}{c|}{\it Linear Coasting with screening}\\
\hline
Temp($ T_9 $) & $ ^4He $ & Temp($ T_9 $) & $ ^4He $ \\[0.5ex]
\hline\hline
$1.000E+02$ & $4.000E-25$  & $1.000E+02$ & $4.000E-25$\\
\hline
$1.000E+02$ & $4.000E-25$  & $1.000E+02$ & $4.000E-25$\\
\hline
$1.000E+02$ & $4.000E-25$  & $1.000E+02$ & $4.000E-25$\\
\hline
$1.000E+02$ & $4.000E-25$  & $1.000E+02$ & $4.000E-25$\\
\hline
$9.999E+01$ & $4.000E-25$  & $9.999E+01$ & $4.000E-25$\\
\hline
$9.876E+01$ & $4.000E-25$  & $9.869E+01$ & $4.000E-25$\\
\hline
$6.590E+01$ & $4.000E-25$  & $6.179E+01$ & $4.000E-25$\\
\hline
$3.148E+01$ & $4.000E-25$ & $2.952E+01$ & $4.000E-25$\\
\hline
$1.504E+01$ & $1.694E-24$ & $1.411E+01$ & $1.327E-24$\\
\hline
$7.645E+00$ & $1.182E-19$ & $7.223E+00$ & $1.439E-19$\\
\hline
$5.710E+00$ & $9.857E-15$ & $5.397E+00$ & $1.188E-14$\\
\hline
$4.766E+00$ & $1.375E-10$ & $4.529E+00$ & $1.562E-10$\\
\hline
$3.783E+00$ & $2.156E-07$ & $3.633E+00$ & $2.436E-08$\\
\hline
$3.693E+00$ & $2.982E-07$ & $3.387E+00$ & $6.010E-08$\\
\hline
$3.066E+00$ & $3.445E-06$ & $2.749E+00$ & $1.202E-06$\\
\hline
$2.383E+00$ & $2.441E-04$ & $2.355E+00$ & $2.095E-05$\\
\hline
$1.790E+00$ & $2.813E-02$ & $2.135E+00$ & $1.798E-04$\\
\hline
$1.495E+00$ & $6.859E-02$ & $1.708E+00$ & $1.980E-02$\\
\hline
$1.279E+00$ & $1.116E-01$ & $1.398E+00$ & $7.419E-02$\\
\hline
$1.091E+00$ & $1.700E-01$ &$1.225E+00$ & $1.195E-01$\\
\hline
$9.273E-01$ & $2.249E-01$ & $1.061E+00$ & $1.806E-01$\\
\hline
$7.622E-01$ & $2.372E-01$ & $8.851E-01$ & $2.324E-01$\\
\hline
$5.835E-01$ & $2.376E-01$ & $7.237E-01$ & $2.384E-01$\\
\hline
$4.940E-01$ & $2.376E-01$ & $5.359E-01$ & $2.386E-01$\\
\hline
$3.637E-01$ & $2.378E-01$ & $4.621E-01$ & $2.386E-01$\\
\hline
$2.360E-01$ & $2.378E-01$ & $3.446E-01$ & $2.386E-01$ \\
\hline
$1.560E-01$ & $2.378E-01$ & $2.275E-01$ & $2.386E-01$\\
\hline
$9.085E-02$ & $2.378E-01$ & $1.480E-01$ & $2.386E-01$\\
\hline
$4.694E-02$ & $2.378E-01$ & $7.696E-02$ & $2.386E-01$\\
\hline
$2.313E-02$ & $2.378E-01$ & $3.885E-02$ & $2.386E-01$\\
\hline
$1.105E-02$ & $2.378E-01$ & $1.874E-02$ & $2.386E-01$\\
\hline
\end{tabular}\label{table}
\end{center}
\end{table} 



\end{document}